\documentclass{aa}
\newcommand{\mdens}{{\rm g~cm^{-3}}}
\newcommand{\bdens}{{\rm fm^{-3}}}
\newcommand{\msun}{{\rm M}_\odot}

\usepackage{graphicx}
\usepackage{amssymb,amsmath}
\usepackage[comma]{natbib}
\usepackage{color}
\usepackage{bm}
\begin{document}
\title{Maximum mass of neutron stars and strange neutron-star cores}
\author{J.L. Zdunik  \and P. Haensel}
\institute{N. Copernicus
Astronomical Center, Polish Academy of Sciences, Bartycka 18,
PL-00-716 Warszawa, Poland
\\{\tt jlz@camk.edu.pl, haensel@camk.edu.pl}}
\offprints{J.L. Zdunik}
\date{Received xxx Accepted xxx}
\abstract{Recent measurement of mass of PSR J1614-2230 rules out most
of existing models of equation of state (EOS) of dense matter with high-density
softening due to hyperonization, based on the recent hyperon-nucleon and
hyperon-hyperon interactions, leading to a "hyperon puzzle".}
{We study a specific  solution of "hyperon puzzle", consisting in replacing a too
soft hyperon core by a sufficiently stiff quark core. In terms of the quark
structure of the matter, one replaces a strangeness carrying baryon  phase of
confined quark triplets, some of them involving s quarks, by  a quark plasma of
deconfined  u, d, and s quarks.}
{We construct an analytic approximation  fitting very well
 modern EOSs of 2SC and CFL color superconducting phases of
 quark matter. Then, we use it to generate  a
 continuum of EOSs of quark matter. This allows us for simulating
 continua of sequences of first-order phase transitions, at prescribed pressures,
  from hadronic matter to the 2SC, and then to the CFL state of color superconducting
  quark matter.  }
{We obtain constraints in the parameter space of the EOS of superconducting quark
cores, EOS.Q,  resulting from $M_{\rm max}>2~\msun$. These constraints depend on
the assumed EOS of baryon phase, EOS.B. We also derive constraints that would
result  from  significantly higher measured masses. For $2.4~\msun$ required
stiffness of the CFL quark core should have been  close to the causality limit, the
density jump at the phase transition being very small. }
{Condition $M_{\rm max}>2~\msun$ puts strong constraints on the EOSs of the 2SC and
CFL phases of quark matter. Density jumps at the phase transitions have to be
sufficiently small and sound speeds in quark matter - sufficiently large. A strict
condition of thermodynamic stability of quark phase results in the maximum mass of
hybrid stars similar to that of purely baryon stars. Therefore, to get $M_{\rm
max}>2~\msun$ for stable hybrid stars, both sufficiently strong additional hyperon
repulsion at high density baryon matter $and$ a sufficiently stiff EOS of quark
matter would be needed. However, it is likely that the high density instability of
quark matter (reconfinement of quark matter) indicates actually the inadequacy of
the point-particle model of baryons in dense matter at $\rho \gtrsim 5\div 8
\rho_0$.}

\keywords{dense matter -- equation of state -- stars: neutron}

\titlerunning{$2\;{\rm M}_\odot$ pulsar and strange cores}
\authorrunning{}
\maketitle

\section{Introduction}
\label{sect:introduction}

The mass of PSR J1614-2230, $1.97\pm 0.04~{\rm M}_\odot$
\citep{Demorest2010}, puts a constraint on the equation of state (EOS)
 of dense matter in neutron star (NS) cores. Namely, maximum allowable mass
 calculated using an acceptable EOS, $M_{\rm max}({\rm EOS})$, should
 be greater than $2.0~\msun$. This proves crucial importance of
 strong interactions in NS cores: their repulsive effect triples
 the value of $M_{\rm max}$ compared to that obtained for non-interacting
 Fermi gas of neutrons, $0.7~\msun$.

 Observational constraint $M_{\rm max}({\rm EOS})>2.0~\msun$
 is easy to satisfy if neutron star cores contain nucleons only,
 and realistic nuclear forces are used, and we get
 $M^{\rm (N)}_{\rm max}({\rm EOS})>2.0~\msun$ for many realistic
 nucleon interaction models \citep{Lattimer2011}. However, nuclear interaction
 models, consistent with experimental data on  hypernuclei,
 predict the presence   of hyperons at the densities
 exceeding $2-3\rho_0$, where normal nuclear density
 $\rho_0=2.7\times 10^{14}~\mdens$  (baryon number density $n_0=0.16~\bdens$).
 Hyperonization of the matter implies a softening of the EOS, due
 to replacing of most energetic neutrons by massive, slowly moving
 hyperons. For realistic models of baryon interactions one gets
 then  $M^{\rm (B)}_{\rm max}\lesssim 1.5~\msun$ (see, e.g.
\citealt{Burgio2011,Vidana2011,SchulzeRijken2011}, and references therein). Such a
low $M_{\rm max}$ is only marginally consistent with $1.44~\msun$ of the
Hulse-Taylor pulsar, but was contradicted already  by  $1.67\pm 0.04~\msun$ of PSR
J1903-0327 (\citealt{Champion2008}; more precise value has been recently obtained
by \citealt{Freire2011}).

Two solutions of the problem of a too low  $M^{\rm (B)}_{\rm max}$ have been
proposed after the discovery of a $2~\msun$ pulsar.
\vskip 2mm
\parindent 0pt
{\bf  Strong hyperon repulsion at high density.} It has been suggested that adding
a new component to the hyperon-hyperon interaction, important for $\rho\gtrsim
5\rho_0$, can stiffen the high-density EOS.B sufficiently to yield $M_{\rm
max}>2.0~\msun$. Repulsive interaction between baryons is supplied by the exchange
of  {\it vector mesons} (spin=1). Hyperon repulsion due to exchange of vector
$\phi$ mesons allows for $M_{\rm max}>2.0~\msun$, without spoiling the agreement
with nuclear and hyper-nuclear data
\citep{Bednarek2011,Weissenborn2011UH,Weissenborn2011phi,Lastowiecki2012}.
 \citet{Dexheimer2008} give an earlier general discussion of vector-meson
contribution to  EOS. An additional increase of $M_{\rm max}$ (above $2.1-2.2~\msun$)
 can be obtained via some breaking of the SU(6) symmetry, usually applied
 to generate vector-meson - hyperon coupling constants
 from the nucleon one \citep{Weissenborn2011phi}. Hyperon repulsion at high
 density is limited by the condition of thermodynamic stability \citep{Bednarek2011}.
\vskip 2mm
\parindent 0pt
{\bf Stiff quark cores in NS.} From the point of view of quantum chromodynamics (QCD),
 appearance of hyperons in dense matter is associated with presence of the  s-quarks,
 in addition to the u and d ones confined into nucleons. Some authors suggested that the
the hyperon core in NS could actually be replaced by  a core of the   u-d-s quark
matter (\citealt{Baldo2003HQ} and  references in \citealt{SchulzeRijken2011}). Let us denote the EOS
of the u-d-s quark matter by EOS.Q. To yield $M_{\rm max}>2.0~\msun$~, quark matter
should have two important (necessary) features: (1) a significant {\it overall
quark repulsion} resulting in a stiff EOS.Q; (2) a strong attraction in a
particular channel resulting in a strong color superconductivity, needed to make
the deconfined Q-phase energetically preferred over the confined B(baryon) one.
After the announcement of the discovery of a $2~\msun$ pulsar, several models of
quark cores of NS (hybrid stars), having the properties necessary to yield $M_{\rm
max}>2.0~\msun$, have been proposed
(\citealt{Ozel2010,Weissenborn2011-MIT,Klaehn2011,Bonanno2012,Lastowiecki2012}).
The EOS of the
hybrid baryon-quark (BQ) stars (EOS.BQ) was constructed using a  two-phase model of
B$\longrightarrow$Q transition, with different underlining theories of the B and Q
phases. \vskip 2mm
\parindent 21pt
There exist many models of color superconducting quark matter states (see, e.g.,
 \citealt{Alford2008}). Two basic states are: two-flavor color superconducting
 (2SC) state and color flavor locked (CFL) superconducting state. In the 2SC
 state only light u and d quarks are paired. The 2SC state is predicted to be
 the ground state of quark matter at $n_{\rm b}\la 4n_0$. On the other hand,
 the CFL superconductor, in which all three flavors are paired, is predicted
 to prevail at high density, $n_{\rm b}\ga 4n_0$. Other superconducting states
 are also predicted \citep{Alford2008}, but they will not be considered here.

In the present paper we derive constraints on the EOS.BQ using an analytical
description of EOS.Q. This allows us to consider a continuum of the EOS.Q models,
or, for a given EOS.B, a continuum of the EOS.BQ models. As the other authors, we
use a two-phase description of 1st order phase transitions (no mixed-phase state;.
neglecting possibility of a mixed-phase state does not influence much the value of
$M_{\rm max}$, see, e.g.,  \citealt{Alford2005}).  We also discuss thermodynamic
stability of the Q phase in stiff quark cores and its impact on $M_{\rm max}$
(this problem has been already mentioned in \citealt{Lastowiecki2012}).

EOSs of baryon matter, used in our work, are described in Sect.\ref{sect:EOS.B}. In
Sect.\ref{sect:Q.continuum} we construct an analytic approximation of the EOS of
quark matter, and we show that it gives a precise approximation of several existing
models of color-superconducting quark cores in neutron stars. Our analytic
approximation is then used to construct a continuum of EOS.Q,
 suitable for constructing an EOS.BQ  with  phase transitions
 at  prescribed pressures, for a given EOS.B.
 In Sect.\ref{sect:BQ.continuum}  we construct continua of the EOS.BQ models, for two
models of EOS.B: a soft one, significantly
 violating a $2.0~\msun$ bound, and a stiff one, satisfying this bound.
 Constraints on EOS.Q, resulting from $M^{\rm (BQ)}_{\rm max}>2~\msun$, are derived in
 Sect.\ref{sect:a.lambda.constraints}.
 We then study, in Sect.\ref{sect:Q.stab}, the  thermodynamic stability of a stiff
 high-density  quark matter. Finally,
 in Sect.\ref{sect:conclusions} we summarize and discuss our results, and point
 out the weak points of our models.

Preliminary results of our work were presented at  the ERPM Pulsar Conference,
Zielona G{\'o}ra, Poland,  24th-27th April, 2012.
\section{EOSs of baryon matter}
\label{sect:EOS.B}
There are two types of existing  EOS.B, leading to  $M_{\rm max}<2~\msun$ and
$M_{\rm max}>2~\msun$, respectively. They will be hereafter referred to as {\it
soft baryon EOS} and {\it stiff baryon EOS}. We will select one EOS belonging to
each of these two groups, and use them to illustrate the features characteristic of
the two types of EOS.B.
\subsection{Soft baryon EOS}
\label{sect:EOS.B.soft}
Most of existing EOS.B yield $M^{\rm (B)}_{\rm max}$ significantly lower than
$2~\msun$. For them replacing a soft hyperon core by a stiff quark one seems to be
an only way of getting $M_{\rm max}>2~\msun$\footnote{It seems that such a possibility was first considered in
\citep{Baldo2003HQ}; we are grateful to David Blaschke for calling our
attention to this paper.}.
 As an example, we consider a very
recent   EOS.B of \citet{SchulzeRijken2011}. This EOS was obtained using
Brueckner-Hartree-Fock many-body approach, and a realistic up-to-date baryon
interaction. In the nucleon sector, Argonne ${\rm V_{18}}$ nucleon-nucleon
potential \citep{WiringaStoks1995} was used, supplemented with a phenomenological
three-body force \citep{LiLombardo2008}. The hyperon-nucleon and hyperon-hyperon
potentials were Nijmegen ESC08
(\citealt{RijkenNagelsYamamoto2010a,RijkenNagelsYamamoto2010b}). This EOS.B gives a
low  $M^{\rm (B)}_{\rm max}=1.35~\msun$ (\citealt{SchulzeRijken2011}). It will be
hereafter referred to as SR one.

\subsection{Stiff baryon EOS}
\label{sect:EOS.B.stiff}
The number of EOS.B  satisfying $M_{\rm max}>2~\msun$ started to increase steadily
after the discovery of a $2~\msun$ pulsar. We selected the BM165  EOS.B of
\citet{Bednarek2011}. This EOS was obtained using non-linear relativistic mean
field model involving the baryon octet coupled to meson fields. The effective
lagrangian includes, in addition to scalar and vector-meson terms, also terms
involving hidden-strangeness scalar and vector-meson coupled to hyperons only.
 For this EOS.B we get  $M^{\rm (B)}_{\rm max}=2.04~\msun$. It will be hereafter
 referred to as BM165.

\section{Quark-matter cores and their EOS}
\label{sect:Q.continuum}
We consider electrically neutral u-d-s quark matter in beta equilibrium and at
$T=0$. The baryon number density  $n_{\rm b}=\frac{1}{3}(n_u+n_d+n_s)$, the energy
density is denoted ${\cal E}$, matter density $\rho={\cal E}/c^2$, and baryon
chemical potential $\mu_{\rm b}={\rm d}{\cal E}/{\rm d}n_{\rm b}$. An important
relation is $n_{\rm b}={\rm d}P/{\rm d}\mu_{\rm b}$. The phase transitions under
consideration are  assumed to be 1st order ones. Therefore, they occur at a
specific (sharp)  value of pressure, and are accompanied by a density jump, from
$\rho_1$ to $\rho_2$.  This is a good approximation for B$\longrightarrow$Q phase
transition in cold dense matter, because the smoothing effect of the mixed B-Q
state is small and can be neglected (\citealt{Endo2006} and references therein). An
important parameter characterizing the density jump at the interface between the
two phases is $\lambda=\rho_2/\rho_1$. Actually, as we have  already mentioned in
Sect.\ref{sect:introduction}, inclusion of a mixed-phase state does not change, to
a very good approximation,  the value of $M_{\rm max}$ \citep{Alford2005}.

\begin{figure}[h]
\resizebox{\columnwidth}{!} {\includegraphics[angle=0,clip]{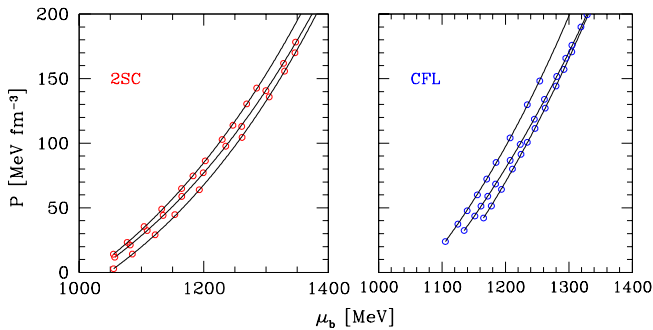}}
\caption{(Online color) Calculated points of EOS of color superconducting quark
matter in the $P-\mu_{\rm b}$ plane \citep{Agrawal2010}  and their approximation by
our analytical formula. See the text for additional details.}
 \label{fig:PmuA}
\end{figure}
\begin{figure}[h]
\resizebox{\columnwidth}{!} {\includegraphics[angle=0,clip]{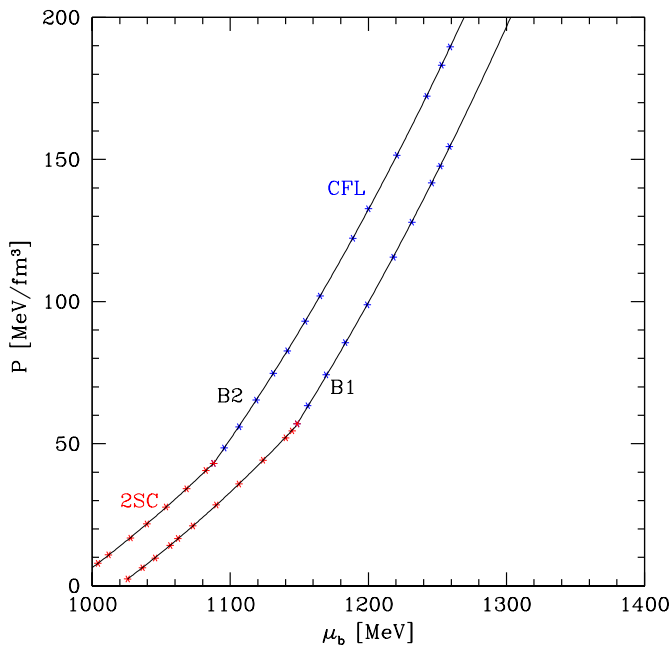}}
\caption{(Online color) Calculated points of EOS of color superconducting quark
matter in the $P-\mu_{\rm b}$ plane \citep{Blaschke2010}  and their approximation
by our analytical formula.
 See the text for additional details.}
 \label{fig:PmuB}
\end{figure}
\subsection{Analytical approximation - 2SC and CFL phases}
\label{sect:analyt.2SC}
 Our method is based on observation that starting from a simple linear formula
 one is able to get a rather precise
 analytic representation of modern EOS.Q(${\cal S}$) in  a phase ${\cal
 S}=$2SC,CFL,
 of color superconducting quark matter under conditions prevailing in neutron
 star cores.  In what follows, we will  usually omit, for simplicity,
  the phase label ${\cal S}$.

Linear EOS ($P$ being a linear function of $\rho$) is characteristic
of a  simplest bag model of quark matter that assumes massless quarks, but it
also holds with  very high accuracy for more realistic  bag model with
massive s-quark \citep{Zdunik2000}.

Linear EOS is determined  by three parameters: $a$, ${\cal E}_\ast$ and $n_\ast$,
where $a$ is square of sound velocity in the units of $c$,
 and  ${\cal E}_\ast$ and $n_\ast$ are energy and baryon
number density at zero pressure, respectively.
We get then:
\begin{equation}
\begin{array}{lcl}
P({\cal E})&=&a ({\cal E}-{\cal E}_\ast)\\
\mu(P)&=&\mu_\ast\cdot\left[1+{1+a\over
a}{P\over{\cal E}_\ast}\right]^{a/(1+a)}\\
n(P)&=&n_\ast\cdot\left[1+{1+a\over
a}{P\over{\cal E}_\ast}\right]^{1/(1+a)}=n_\ast\left(\frac{\mu}{\mu_\ast}\right)^{1/a}\end{array}
\label{eq:lin}
\end{equation}
where $\mu_\ast={\cal E}_\ast/n_\ast$ is baryon chemical potential at zero pressure.
The stiffness of the matter is described by the parameter
$a={\rm d}P/{\rm d}{\cal
E}=(v_{\rm sound}/c)^2$. Special cases of linear EOSs  with $a=1$ and $a=1/3$
were recently considered  by \cite{Chamel2012} in their study of exotic cores in
neutron stars.

Numerical results for a quark matter  EOS are usually given as points in the
$P-\mu_{\rm b}$ plane \citep{Agrawal2010,Blaschke2010}.
These varables are very convenient to study microscopic stability of
matter and for determination of phase transition,
which correspond to the crossing point of $P(\mu)$ relations for different phases.
The density jump at phase transition is described then by the change of the slope
of $P(\mu)$ function through the relations:
$n=dP/d\mu$ and ${\cal E}=n\mu-P$ (see, e.g.,  Fig \ref{fig:PmuB})

We assume that the linear  EOS, Eq.(\ref{eq:lin}),  accurately describes
quark matter cores in neutron stars, corresponding to baryon density range
 $2n_0\la n_{\rm b}\la 10 n_0$  Let us stress, that the use of linear dependence,
Eq.(\ref{eq:lin}), is restricted to  $2n_0-10n_0$ (or $300~{\rm MeV}~\bdens
<{\cal E}<1500~{\rm MeV}~\bdens$, or  $30~{\rm MeV}~\bdens <P<300~{\rm
MeV}~\bdens$),  and by no means is claimed to be valid outside the neutron-star
core regime.

Let us introduce dimensionless quantities: $\overline{n}_{\rm b}\equiv n_{\rm b}/n_\ast$ and
$\overline{\mu}_{\rm b}\equiv \mu_{\rm b}/\mu_\ast$. We can then use
Eq.(\ref{eq:lin}) to get $P$ as a function of
$\overline{\mu}_{\rm b}$,
\begin{equation}
{P}={{\cal E}_\ast\over \nu}
(\overline{\mu}_{\rm b}^\nu -1)~,
\label{eq:overline.P.mu.lin}
\end{equation}
where $\nu\equiv (1+a)/a$.
Then an analytical approximation for $\overline{n}_{\rm b}$
reads
\begin{equation}
\overline{n}={{\rm d}P(\mu_{\rm b})\over {\rm d}\mu_{\rm b}}
=\overline{\mu}_{\rm b}^{\nu-1}~.
\label{ref:overline.n.mub}
\end{equation}

Determination of the value of $n_\ast$ deserves an additional comment.
 It can be taken from original numerical calculations if available.
 In not directly available, it can be calculated from the original plot
 of $P^{\rm (calc)}(\mu_{\rm b})$ using
 $n_\ast=\left({\rm d}P^{\rm (calc)}/{\rm d}\mu_{\rm b}\right)_{\mu_\ast}$.

We now pass to specific cases of ${\cal S}$=2SC, CFL.
The  least-squares fit method results in curves presented in Figs.
\ref{fig:PmuA}, \ref{fig:PmuB}. This fit works very well and could be also
checked by comparing values of $n=dP/d\mu$ with exact results (if available,
like in \citealt{Agrawal2010}).

\vskip 2mm
\parindent 0pt
{\bf 2SC} In view of its intermediate-density range, $2n_0 \la n_{\rm b}\la 4n_0$,
the 2SC state is less important for $M_{\rm max}$ than the high-density  CFL state
realized for $n_{\rm b}>4n_0$. However, as we will show, the softening due to the
density jump at the B-Q(2SC) transition has a significant  indirect effect on the
$2~\msun$ constraint imposed on the EOS of the CFL phase. We considered two
numerical EOS.Q(2SC)s, calculated by \cite{Agrawal2010} and two of
\cite{Blaschke2010}.  All these EOSs were calculated using the Nambu - Jona-Lasinio
(NJL) model of quark matter and color superconductivity. The NJL model is a
non-perturbative low-energy approximation to QCD. As seen in Figures
\ref{fig:PmuA}, \ref{fig:PmuB}, our analytic formulae fit numerical results very
precisely. It should be stressed, that these analytical formulae reproduce also very
well numerically calculated points in the $P-n_{\rm b}$ plane, whenever these
points are available, e.g., in \cite{Agrawal2010}. Our approximation in this case
gives the value of parameter $\nu =4.1\div 4.6$ which corresponds to
$a=0.25\div0.33$.

\vskip 2mm {\bf CFL} The baryon density interval $4n_0\la n_{\rm b}\la 10n_0$ is
crucial for the value of $M_{\rm max}$. Therefore, it is the EOS in the CFL state
which is decisive for the value of  $M^{\rm (BQ)}_{\rm max}$. We considered five
numerical EOS of CFL  superconducting quark matter, three models from
\citep{Agrawal2010} and two models from \citep{Blaschke2010}. All of them were
based on the NJL model.
 As we see in Figures \ref{fig:PmuA}, \ref{fig:PmuB}
our analytical formulae  are very precise. Similarly as for the 2SC phase,  these
formulae reproduce also very well numerically calculated points in the
$P-n_{\rm b}$ plane. The CFL phase is stiffer than the 2SC one: the  values of $a$
range within 0.3 and 0.4 ($\nu=3.5\div 4.3$).
\parindent 21pt
\subsection{A family of analytical  models of EOS.BQ}
\label{sect:BQ.continuum}
We generalize now  discrete sets $\lbrace {\cal E}_{_{\cal S}\ast}^i,a_{_{\cal S}}^i,
n_{_{\cal S}\ast}^i\rbrace$ into  a continuum of three-parameter  models $\lbrace
{\cal E}_{_{\cal S}\ast}, a_{_{\cal S}},n_{_{\cal S}\ast}\rbrace$, within a region of
parameter space determined by appropriate constraints on these  parameters. We
assume that $0.2< a <0.8$, while  $2n_0<n<10n_0$ and $10~{\rm
~MeV}~\bdens<P<300~{\rm MeV}~\bdens$

Having constructed a continuum of the EOS.Q(${\cal S}$) models we are able to
simulate, for a given EOS.B, a sequence of  phase transition
B$\longrightarrow$Q(2SC)$\longrightarrow$Q(CFL). \vskip 2mm
\parindent 0pt
{\bf Transition from B to Q(2SC)} Assume that  B$\longrightarrow$Q(2SC) takes place
at $P=P_1$. Three parameters of  an  EOS.Q(2SC) taken from our family are then
interrelated by two conditions at the B$\longrightarrow$Q(2SC) phase transition
point : continuity of the baryon chemical potential, and continuity of the
pressure,
\begin{equation}
\mu_{\rm b}^{{\rm (B)}}(P_1)= \mu_{\rm b}^{{\rm ({2SC})}}(P_1)~, ~~ P^{{\rm
(B)}}({\cal E}_1)=P^{{\rm (2SC)}}({\cal E}_2)~. \label{eq:B.Q.eq}
\end{equation}
 Upper indices (B) and (2SC) refer to  the baryon phase,  and the 2SC quark phase,
 respectively.  Now, let us fix the EOS   of baryon matter, EOS.B.  Models of the phase
 transition  to the 2SC quark matter  are then labeled by
 $P_1=P^{\rm (B)}({\cal E}_1)$  and the relative density jump
 $\lambda_{_{\rm 2SC}}={\cal E}_2/{\cal
 E}_1$,  with thermodynamical parameters ($n_1,n_2,{\cal E}_1,{\cal E}_2$) satisfying
Equations (\ref{eq:B.Q.eq}). Here, the index 1 refers to the B phase, and 2 to the
2SC phase.  \vskip 2mm

{\bf Transition from Q(2SC) to Q(CFL)}
 Let us now choose  the pressure at which   the 2SC$\longrightarrow$CFL
 transition occurs, $P=P_2$. Using conditions of  continuity of the pressure and
 of the baryon chemical potential at $P=P_2$, we obtain a one-parameter family
 $\lbrace {\rm EOS.Q(CFL)}\rbrace$  attached to  a specific EOS.Q(2SC)
 from the previously constructed $\lbrace {\rm EOS.BQ(2SC)}\rbrace$ family.
 A continuous parameter  within  $\lbrace {\rm EOS.Q(CFL)}\rbrace$ can
 be   $a_{_{\rm CFL}}$ or  $\lambda_{_{\rm CFL}}$. This completes
 the second step in the procedure of constructing  a general
  family $\lbrace {\rm EOS.BQ}\rbrace$ with
 B$\longrightarrow$Q(2SC)$\longrightarrow$Q(CFL) phase transitions
 at prescribed pressures  $P_1$ and $P_2$, respectively.
 \vskip 2mm

 In order to have a "reference one-phase quark core" we will also we
 consider EOS.BQ(CFL),  with a phase transition from B directly to the
 CFL phase of quark matter, at pressure $P_{\rm _{CFL}}$ and density
 $\rho_{\rm _{CFL}}$.
\vskip 2mm \parindent 0 pt

\begin{figure}[h]
\resizebox{\columnwidth}{!} {\includegraphics[angle=0,clip]{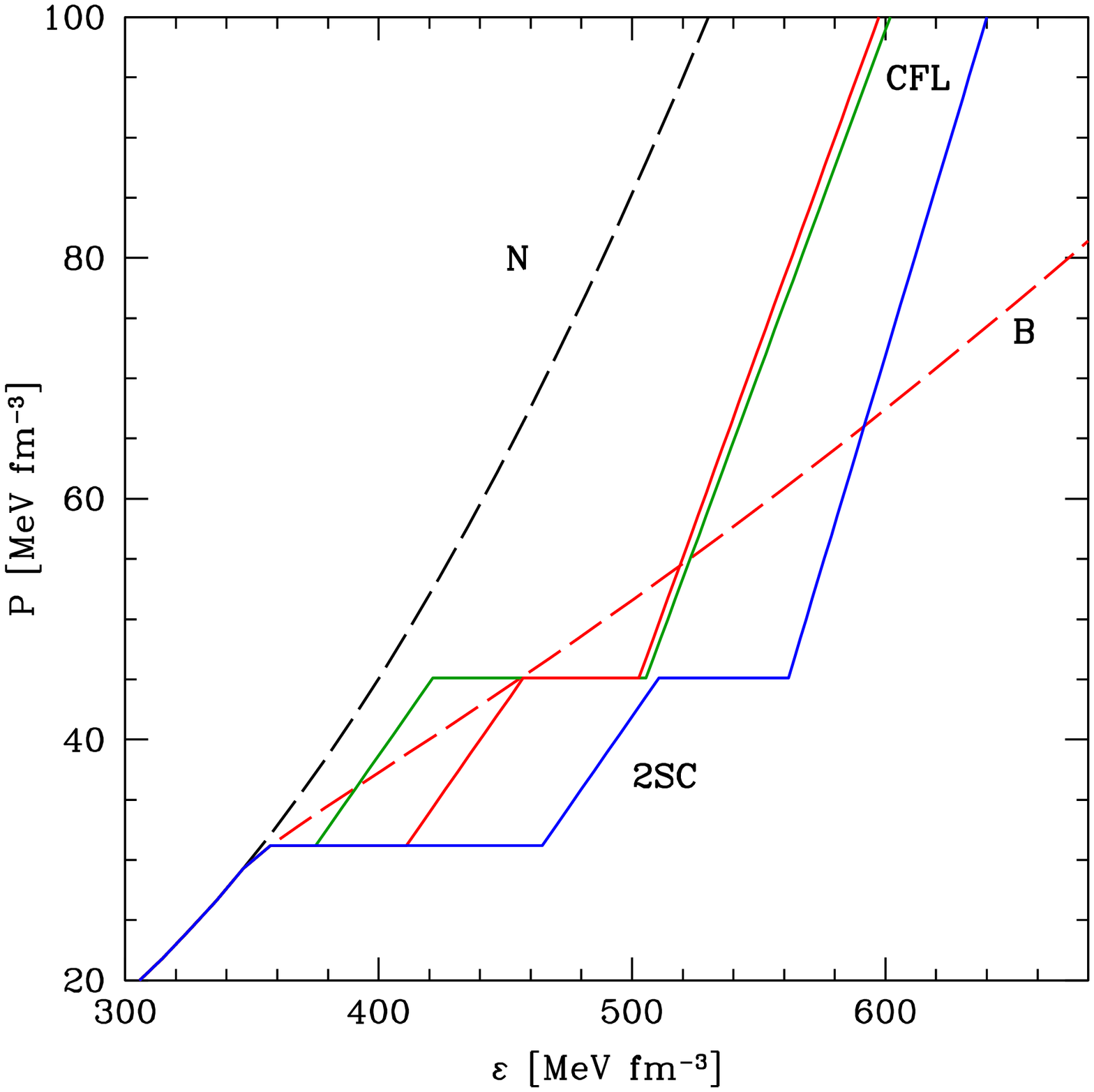}}
\caption{(Online color) Examples of EOS.BQ from a continuum $\lbrace{{\rm
EOS.BQ}}\rbrace$ emerging from a soft SR  EOS. Hadronic EOSs are plotted as dashed
lines: N - nucleon EOS; B - nucleons and hyperons. The phase transition
B$\longrightarrow$Q(2SC) takes place at $P_1=31~{\rm MeV~}\bdens$, and
Q(2SC)$\longrightarrow$Q(CFL) at $P_2=45~{\rm MeV~}\bdens$. The  2SC phase has
$a_{_{\rm 2SC}}=0.302$. Three examples of
B$\longrightarrow$Q(2SC)$\longrightarrow$Q(CFL) are shown corresponding to
following choices of $\lbrace \lambda_{_{\rm 2SC}}, \lambda_{_{\rm CFL}},a_{_{\rm
CFL}}\rbrace$: $\lbrace 1.05,1.2,0.57\rbrace$, $\lbrace
1.15,1.1,0.58\rbrace$,$\lbrace 1.3,1.1,0.7\rbrace$. In all these
 cases maximum mass is equal to $2\msun$, i.e., parameters  
$\lbrace \lambda_{_{\rm 2SC}}, \lambda_{_{\rm CFL}},a_{_{\rm CFL}}\rbrace$
 lie on the bounding curves in Fig.
\ref{fig:lamaHJS}.}
 \label{fig:PE1}
\end{figure}

 {\bf  Direct transition from B to Q(CFL) at $P_{\rm _{CFL}}$}. For
$P_1<P_{\rm _{CFL}}<P_2$ this EOS can be either softer or stiffer than in the case
of B$\longrightarrow$2SC$\longrightarrow$CFL, depending on the softness of the
hyperonic EOS. This is illustrated in Fig. \ref{fig:PE1} where the ``middle'' model
of B$\longrightarrow$2SC transition (with $\lambda_{_{\rm 2SC}}=1.15$) gives a mean
stiffness similar to that of the hyperon (B) phase (i.e. pressures and densities at
the bottom of CFL core are almost the same in both cases).

 \parindent 21pt
 \vskip 2mm

\begin{figure}[h]
\resizebox{\columnwidth}{!} {\includegraphics[angle=0,clip]{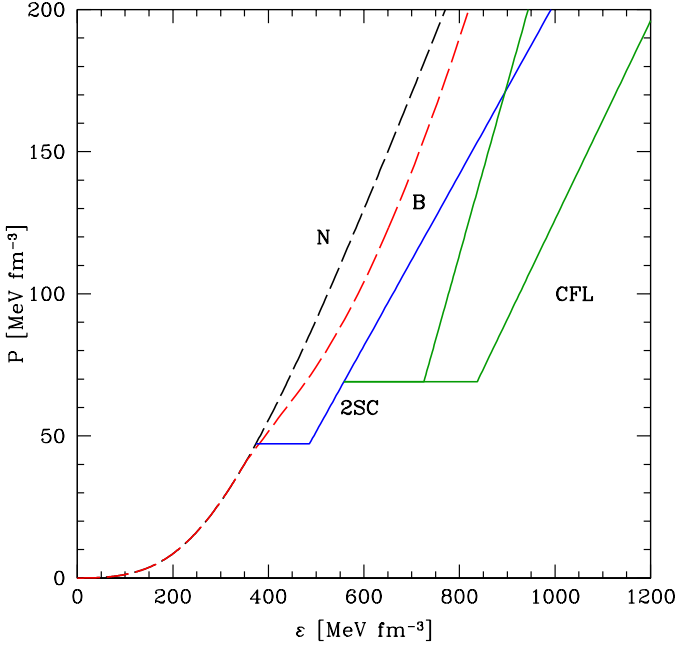}}
\caption{(Online color) Examples of EOS.BQ from a continuum $\lbrace{{\rm
EOS.BQ}}\rbrace$ emerging from a stiff MB165 EOS. Hadronic EOSs are plotted as
dashed lines: N - nucleon EOS (black); B - nucleons and hyperons (red). The phase
transition B$\longrightarrow$Q(2SC) takes place at $P_1=47~{\rm MeV~}\bdens$,
$a_{_{\rm 2SC}}=0.302$ and  $\lambda_{_{\rm 2SC}}=1.3$. EOS.Q(2SC) is plotted as a
blue line. Two examples of Q(2SC)$\longrightarrow$Q(CFL) at $P_2=69~{\rm
MeV~}\bdens$ are shown, corresponding to a stiffer CFL phase ($a_{_{\rm CFL}}=0.6$,
$\lambda_{_{\rm 2SC}}=1.3$) and a softer one ($a_{_{\rm CFL}}=0.35$,
$\lambda_{_{\rm 2SC}}=1.5$). The EOS.Q(CFL)s are plotted as green solid lines.}
 \label{fig:PE2}
\end{figure}
\parindent 21pt
 Several examples of  EOS.BQ constructed following the procedure described
 above  are shown in  Figures \ref{fig:PE1}, \ref{fig:PE2}.

For two considered EOS.B, i.e. SR and BM165,  we are using two different choices of
transition pressures $P_1$ and $P_2$. Here, $P_1$ corresponds to the pressure at which in
original models hyperons start to appear. For SR EOS $P_1\simeq 30~{\rm
MeV~}\bdens$, while for BM165  EOS  $P_1\simeq 50{\rm MeV~}\bdens$. Thickness of
the 2SC phase layer corresponds to  $\Delta P=P_2-P_1\approx 20~{\rm MeV~}\bdens$.
In both cases pressure $P_1$ corresponds to $n_{\rm b}\approx 2 n_0$.
\section{Constraints on EOS.Q in the $a-\lambda$ plane}
\label{sect:a.lambda.constraints}
\begin{figure}[h]
\resizebox{\columnwidth}{!} {\includegraphics[angle=0,clip]{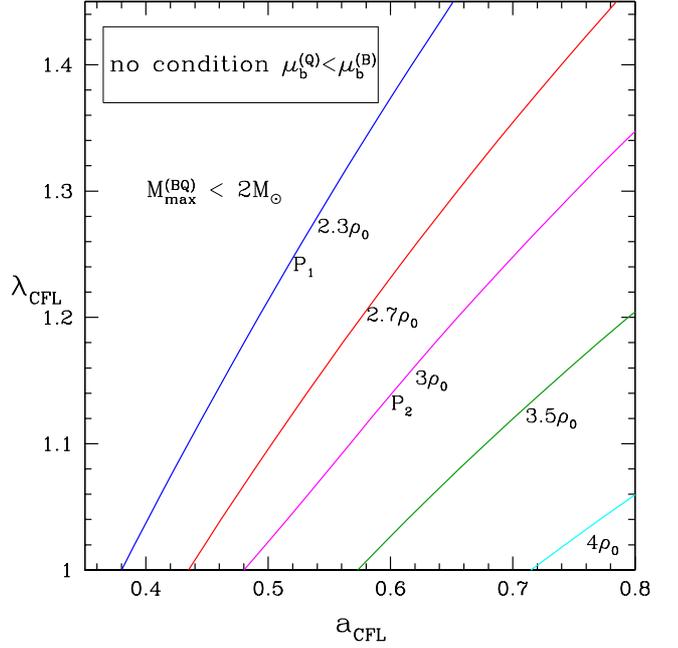}}
\caption{(Color online) Constraints in the $a-\lambda$ plane,  resulting from
$M_{\rm max}>2~\msun$, obtained assuming SR EOS of baryon matter, and for purely
CFL cores (no 2SC layer) starting at $\rho_{\rm _{CFL}}=2.3\rho_0,...,4.0\rho_0$.
Densities $2.3\rho_0$ and $3.0\rho_0$ correspond to $P_1$ and $P_2$ relevant for
the boundaries of the 2SC layer in a 2SC+CFL core.  Each line is an upper boundary
of the region of $(a_{_{\rm CFL}},\lambda_{_{\rm CFL}})$ consistent with $M_{\rm
max}>2~\msun$. These lines are labeled by the density $\rho_{\rm _{CFL}}$ at which
a direct transition B$\longrightarrow$Q(CFL) takes place.}
 \label{fig:lamaCFL}
\end{figure}

\begin{figure}[h]
\resizebox{\columnwidth}{!} {\includegraphics[angle=0,clip]{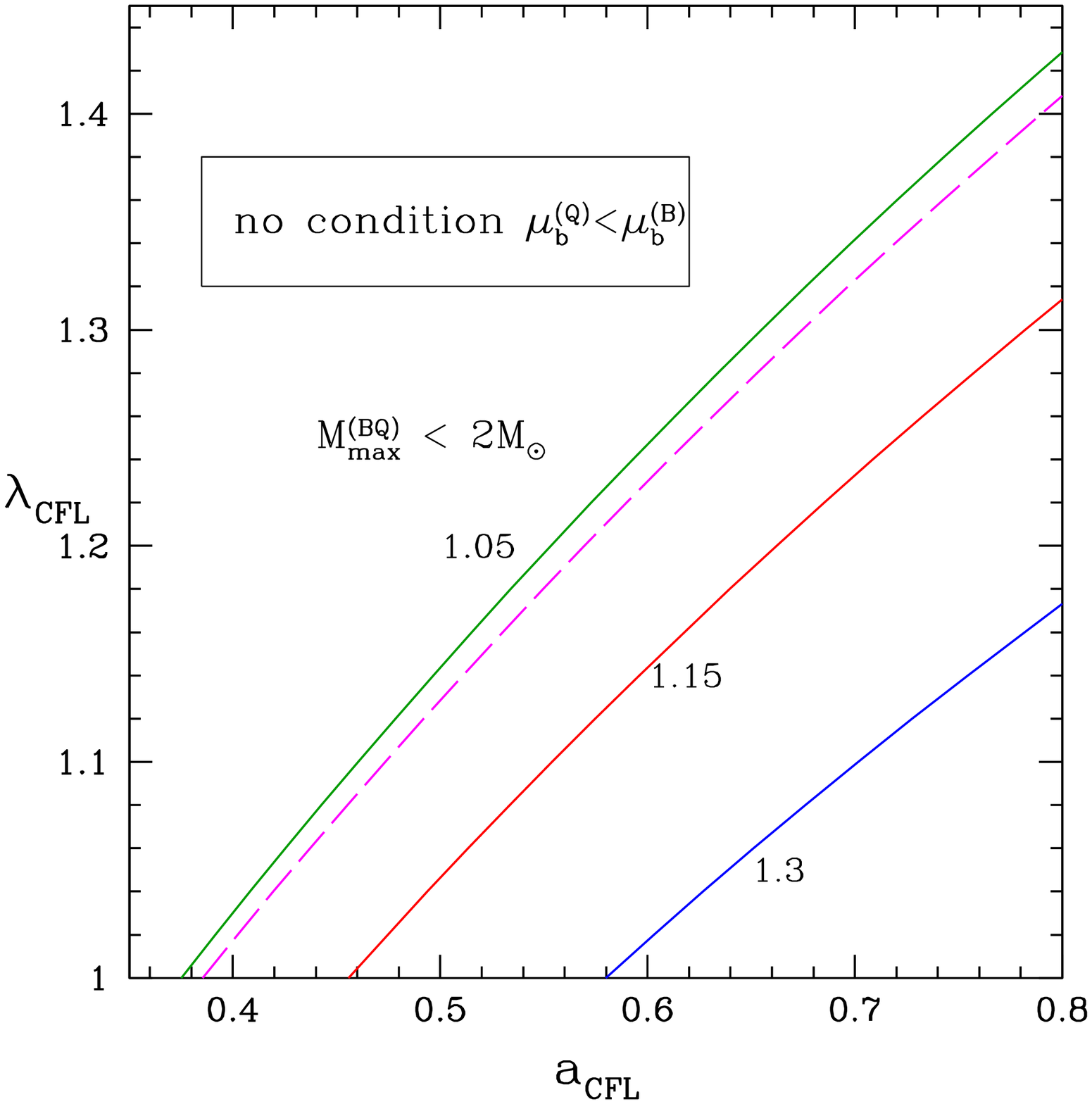}}
\caption{(Color online) Constraints in the $a-\lambda$ plane,  resulting from
$M_{\rm max}>2~\msun$, obtained assuming BM165 EOS of baryon matter, and quark core
starting at $P_1=47~{\rm MeV~fm^{-3}}$. Each line is an upper boundary of the
region of $(a_{_{\rm CFL}},\lambda_{_{\rm CFL}})$ consistent with $M_{\rm
max}>2~\msun$. Solid lines are obtained for quark cores composed of a 2SC layer
$(P_1<P<P_2)$ and a CFL core starting at $P_2=69{\rm MeV~fm^{-3}}$. These lines are
labeled by the density jump at the B-2SC interface $\lambda_{_{\rm 2SC}}=1.05,
1.15, 1.30$.
 We used $a_{_{\rm 2SC}}=0.3$ \citep{Agrawal2010}.
Dashed line is obtained for purely CFL cores (no 2SC layer) starting at $P_2$.
 } \label{fig:lamaBM}
\end{figure}
\begin{figure}[h]
\resizebox{\columnwidth}{!} {\includegraphics[angle=0,clip]{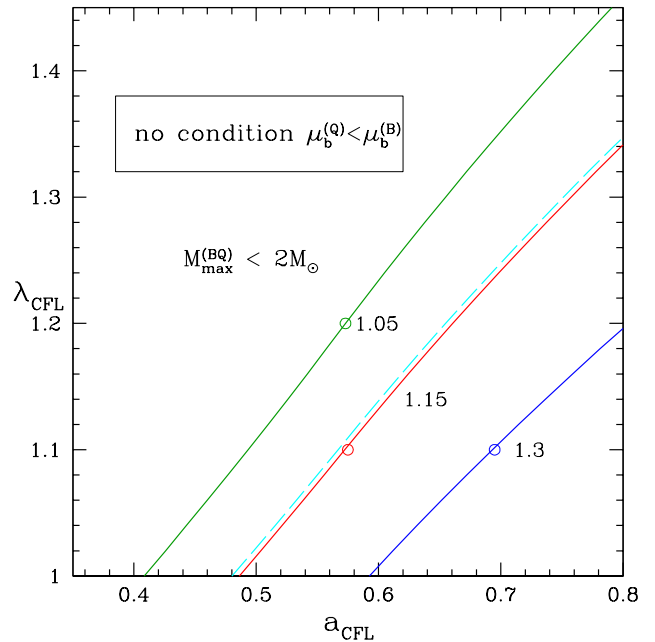}}
\caption{(Color online) Constraints in the $a-\lambda$ plane, resulting from
$M_{\rm max}>2~\msun$, obtained assuming SR EOS of baryon matter, and quark core
starting at $P_1=31~ {\rm MeV~fm^{-3}}$. Each line is an upper boundary of the
region of $(a_{_{\rm CFL}},\lambda_{_{\rm CFL}})$ consistent with $M_{\rm
max}>2~\msun$. Solid lines are obtained for quark cores composed of a 2SC layer
$(P_1<P<P_2)$ and a CFL core starting at $P_2=45~{\rm MeV~fm^{-3}}$.
 These lines are labeled by
the density jump at the B-2SC interface $\lambda_{_{\rm 2SC}}=1.05, 1.15, 1.30$. We
used $a_{_{\rm 2SC}}=0.3$ \citep{Agrawal2010}. Dashed line is obtained for purely
CFL cores (no 2SC layer) starting at $P_2=45~{\rm MeV~fm^{-3}}$.
Open circles correspond to EOSs presented in Fig. \ref{fig:PE1}.
}
\label{fig:lamaHJS}
\end{figure}


A  1st order phase transition in the neutron star core affects the EOS in two ways.
First, it softens it because $\lambda>1$ implies some density range with ${\rm
d}P/{\rm d}\rho =0$. Second, a new phase is either stiffer or softer than the less
dense phase, the stiffness of the quark phase being determined by $a$.  The
condition $M_{\rm max}>2~\msun$ imposes therefore a condition in the $a-\lambda$
plane. In what follows, we start with a case of a purely CFL quark core, and then
we consider the quark core composed of an outer 2SC layer and an inner CFL core.
Our results will be  illustrated by examples presented in Figs \ref{fig:lamaCFL}-\ref{fig:lamaHJS}. As
in the preceding sections, we considered two EOS of baryon matter, a soft SR EOS,
and a stiff BM165 EOS.

\vskip 2mm
\parindent 0pt
{\bf Pure CFL core} We consider EOS.BQ(CFL), with quark core edge at $P=P_{_{\rm
CFL}}$. We calculate a locus of points in the $a_{_{\rm CFL}}-\lambda_{_{\rm CFL}}$
plane corresponding to $M_{\rm max}=2~\msun$
 This locus is a line $\lambda^{_{\rm CFL}}_{\rm max}(a_{_{\rm CFL}})$ such that
 points $(a,\lambda)$ below it generate Q-cores  satisfying $M_{\rm max}>2~\msun$,
 while those above it violate this condition. We can alternatively
 call this locus line  $a^{_{\rm CFL}}_{\rm min}(\lambda_{_{\rm CFL}})$.
Of course the location of lines  $a_{_{\rm CFL}}-\lambda_{_{\rm CFL}}$ depends on
the value of pressure $P_{\rm _{CFL}}$(density $\rho_{\rm _{CFL}}$) at which phase
transition to quark core (CFL) occurs. This dependence is presented in Fig.
\ref{fig:lamaCFL}.

\vskip 2mm
\parindent 0pt
{\bf 2SC+CFL core, effect of a 2SC layer: general procedure}
We replace an outer layer of
the matter in the B phase containing hyperons (or a layer of the  CFL core with
pressures
 $P_{\rm _{CFL}}<P<P_2$), by a  layer of the 2SC phase. The effect of
this additional layer of the 2SC phase depends on the value of $P_{\rm _{CFL}}$. If
$P_{\rm _{CFL}}=P_1$ we replace an outer layer of the CFL core by the quark matter
in the 2SC phase. This will result in a softening of the quark core since 2SC phase
is thought to be softer  than the CFL one (due to a much smaller superfluid gap),
and also because of an  additional density jump $\lambda_{_{\rm 2SC}}$. Even if
$\lambda_{_{\rm 2SC}}$  is very close to one, we get
 $\lambda_{\rm max}^{_{\rm 2SC+CFL}}(a)< \lambda_{\rm max}^{_{\rm
 CFL}}(a)$. However, the effect of 2SC phase on the value of
 $\lambda^{_{\rm 2SC+CFL}}_{\rm max}$  is then rather small. With an increasing
 value of $\lambda_{_{\rm 2SC}}$,  the effect of softening gets stronger, and an allowed
 region of $(a_{_{\rm CFL}},\lambda_{_{\rm CFL}})$ gets smaller.
 In the limit of $P_{\rm CFL}=P_2$ we replace baryon  matter by the 2SC phase
and the net effect depends on the relative stiffness of these two phases (see
Fig.\ref{fig:PE1} and discussion at the end of Sect. \ref{sect:BQ.continuum}).

 \vskip 2mm\parindent 0pt
{\bf Stiff EOS.B, 2SC+CFL core} The loci $a^{_{\rm CFL}}_{\rm min}(\lambda_{_{\rm
CFL}})$ are shown in Fig.\ref{fig:lamaBM}. Although the mass fraction contained in
the 2SC layer is small, its effect on the size of the allowed $(a_{_{\rm
CFL}},\lambda_{_{\rm CFL}})$ region is strong. For $\lambda_{_{\rm CFL}}>1.2$,
required $a_{_{\rm CFL}}$ has to be significantly larger than the values obtained
in \citep{Blaschke2010,Agrawal2010}. Simultaneously,  at $a_{_{\rm CFL}}\approx
0.5$ the density jump due to the 2SC$\longrightarrow$CFL transition is constrained
to values significantly below the ones obtained in
\citep{Blaschke2010,Agrawal2010}.

\vskip 2mm\parindent 0pt {\bf Soft EOS.B, 2SC+CFL core}  The loci
 $a^{_{\rm CFL}}_{\rm min}(\lambda_{_{\rm CFL}})$ are shown in
 Fig.\ref{fig:lamaHJS}. Even for a very low density jump  $\lambda_{_{\rm
 2SC}}=1.05$, we obtain $a_{_{\rm CFL}}>0.4$, which is rather stringent.
 In our case, the  result obtained for B$\longrightarrow$Q(CFL) at $P_{\rm _{CFL}}=P_2$ is very
similar to that of B$\longrightarrow$Q(2SC)$\longrightarrow$Q(CFL) with
$\lambda_{\rm 2SC}=1.15$ (the dashed line is very close to the solid one) .
\parindent 21pt
\begin{figure}[h]
\resizebox{\columnwidth}{!} {\includegraphics[angle=0,clip]{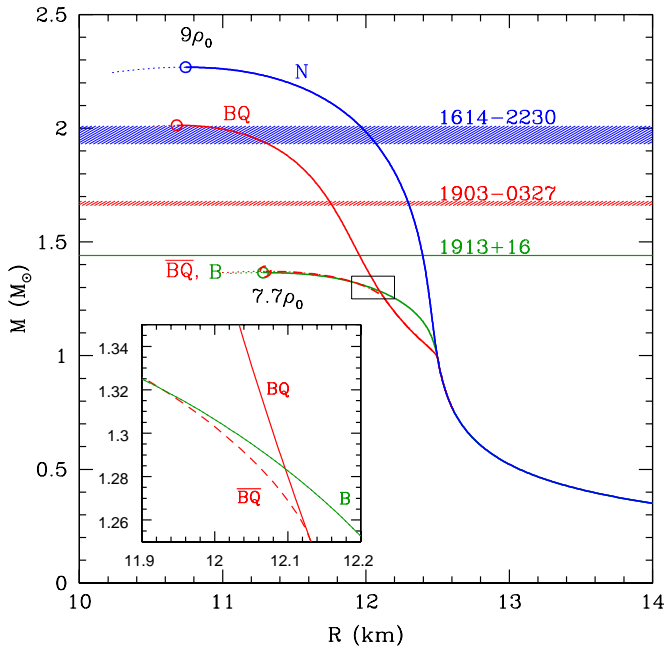}}
\caption{(Color online) Mass $M$ vs. radius $R$ for non-rotating
neutron star models calculated using different assumptions concerning structure of
dense matter for $\rho>2\rho_0$. Dotted segments of $M(R)$
 curves: configurations unstable with respect to small radial perturbations.
 N - nucleon matter, Argonne ${\rm V_{18}}$+TBF nucleon interaction
 \citep{SchulzeRijken2011}. B - baryon matter, EOS calculated using
 hyperon-nucleon and  hyperon-hyperon potentials  Nijmegen ESC08
 (\citealt{RijkenNagelsYamamoto2010a,RijkenNagelsYamamoto2010b}). Open
 circles -  configurations of maximum allowable mass, with central density given in the units
 of $\rho_0$.  BQ - hybrid stars with quark cores,  when condition
 $\mu_{\rm b}^{{\rm (Q)}}(P)<\mu_{\rm b}^{{\rm (B)}}(P)$ is not imposed.
 $\overline{\rm BQ}$ - hybrid stars with stable quark cores, when condition
 $\mu_{\rm b}^{{\rm (Q)}}(P)<\mu_{\rm b}^{{\rm (B)}}(P)$ is satisfied.
 Analytical EOS of quark
 matter in the CFL state with $\rho_2=2.35\rho_0$, $a_{_{\rm CFL}}=0.5$,
 $\lambda_{_{\rm CFL}}=1.2$.
Enlarged rectangle:  Vicinity of the  high-density 1st order phase transition
Q$\longrightarrow$B (reconfinement). } \label{fig:MR_SR}
\end{figure}
\begin{figure}[h]
\resizebox{\columnwidth}{!} {\includegraphics[angle=0,clip]{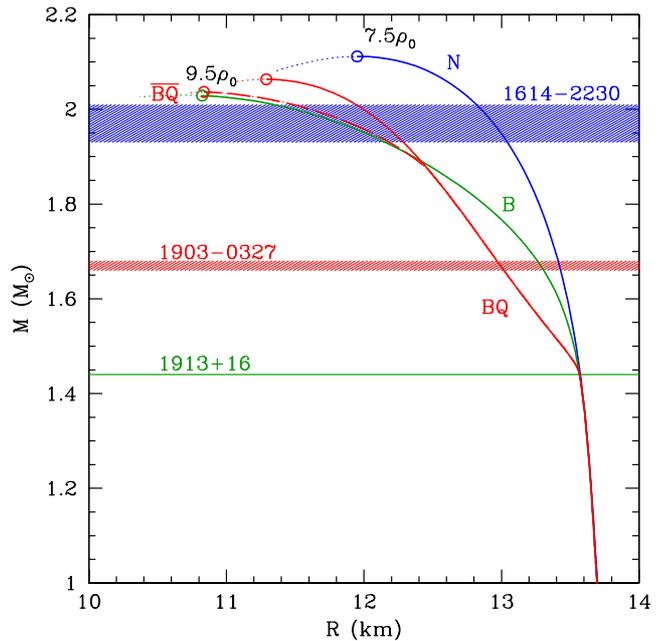}}
\caption{(Color online) Same notations as in Fig.\ref{fig:MR_SR}, but for BM165 EOS
of nucleon  and baryon matter  \citep{Bednarek2011}. Analytical EOS of quark
 matter and EOS.Q(CFL)  with $\rho_2=2.5\rho_0$, $a_{_{\rm CFL}}=0.5$,
 $\lambda_{_{\rm CFL}}=1.2$.}
 \label{fig:MR_BM}
\end{figure}
\section{Stability of quark cores and $M_{\rm max}$}
\label{sect:Q.stab}
Up to now, we did not consider the (thermodynamical) {\it stability} of a stiff
quark core in a hybrid (BQ) star. A stiffening of the EOS is necessarily associated
with the increase of the baryon chemical potential (see an example in
\citealt{Bednarek2011}). In particular, it may lead to the thermodynamical
instability of the stiff (Q) phase with respect to the re-conversion into the (B)
one. This instability  results from the violation, above a certain pressure, of the
condition $\mu_{\rm b}^{{\rm (Q)}}(P)<\mu_{\rm b}^{{\rm (B)}}(P)$. Assuming a
complete thermodynamical equilibrium, we are dealing with a first-order phase
transition back to the (B) phase, that one can call  {\it reconfinement}
(cf., \citealt{Lastowiecki2012}).
A corresponding EOS will be denoted EOS.$\overline{\rm BQ}$ and the $M(R)$ branch
based on this EOS will be labeled $\overline{\rm BQ}$. Examples of the  N,B,BQ and
$\overline{\rm BQ}$ branches in the $M-R$ plane, obtained for a soft SR EOS of
baryon matter are presented in Fig.\ref{fig:MR_SR}. The reconversion
$Q\longrightarrow B$ strongly limits the size of the quark core in hybrid stars and
results in the value of $M^{_{\overline{\rm BQ}}}_{\rm max}\approx M^{\rm (B)}_{\rm
max}=1.35~\msun$ (Fig.\ref{fig:MR_SR})

 For the BM165 EOS.B we get  $M^{\rm (B)}_{\rm max}=2.04~\msun$.
 Replacing hyperon cores  by stiff quark ones can further increase the value
 of $M_{\rm max}$.  An example is shown in Fig.\ref{fig:MR_BM}, where we obtain
  $M^{\rm (BQ)}_{\rm max}=2.07~\msun$.  However, if complete thermodynamic
 equilibrium is imposed, the Q$\longrightarrow$B transition back to
 the B phase takes place and one gets maximum allowable mass
 $M^{_{\overline{\rm BQ}}}_{\rm max}\approx M^{\rm (B)}_{\rm max}=2.04~\msun$.

\section{Summary, discussion,  and conclusions}
\label{sect:conclusions}
The existence of a $2~{\rm M}_\odot$ pulsar is a challenge for neutron star models
with strangeness carrying cores. Strangeness is associated with s quark, either
confined into hyperons or moving in a (deconfined) quark plasma.

 The threshold density for the appearance of hyperons, predicted by realistic models
  of dense matter consistent with nuclear and hypernuclear data, is
 $\sim 2\rho_0-3\rho_0$.  Realistic baryon
interactions lead to $M_{\rm max}$,  for neutron stars with hyperon cores starting
at such density, that is significantly below $M=2.0~\msun$. This contradiction can
 be removed  due a hypothetical strong high-density repulsion acting between
hyperons. As discussed in several papers, this  strong high-density  repulsion
could result from the exchange of a vector meson $\phi$ coupled only to hyperons.
\begin{figure}[h]
\resizebox{\columnwidth}{!} {\includegraphics[angle=0,clip]{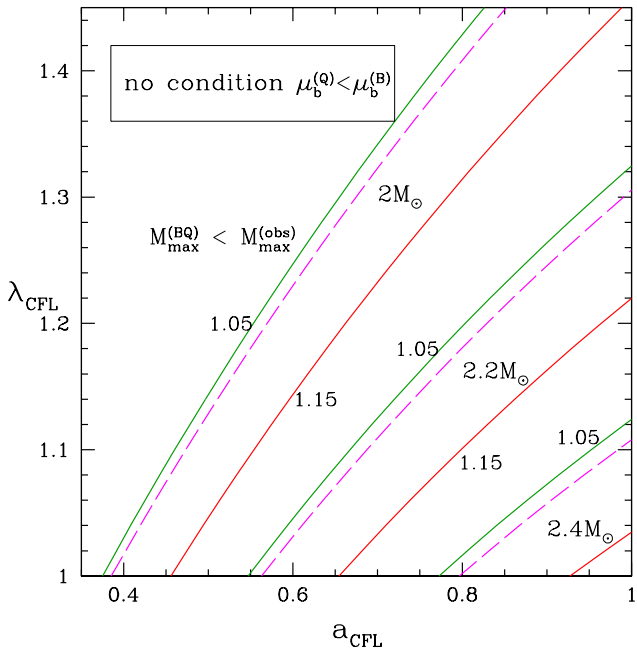}}
\caption{(Color online) Constraints in the $a-\lambda$ plane,
 resulting from 3 different values  of maximum measured mass
$M_{\rm max}^{\rm (obs)}=2.0~\msun, 2.2~\msun, 2.4~\msun$. We assume  BM165 EOS of
baryon matter, and quark core starting at $P_1=47~{\rm MeV~fm^{-3}}$. Each line is
an upper boundary of the region of $(a_{_{\rm CFL}},\lambda_{_{\rm CFL}})$
consistent with $M_{\rm max}>M^{\rm (obs)}_{\rm max}$. Solid lines are  obtained
for purely CFL cores (no 2SC layer). Dashed lines are obtained for quark cores
composed of a 2SC layer $(P_1<P<P_2)$ and a CFL core starting at $P_2=69~{\rm
MeV~fm^{-3}}$. These lines are labeled by the density jump at the B-2SC interface
$\lambda_{_{\rm 2SC}}=1.05, 1.15$. We used $a_{_{\rm 2SC}}=0.3$
\citep{Agrawal2010}. } \label{fig:lama_mass}
\end{figure}

Howewer, it has also been considered that massive neutron stars could actually be
hybrid stars with stiff quark-matter cores that allow for $M>2~\msun$. Strong
overall repulsion between quarks should be accompanied by a strong attraction
(pairing) in a specific two-quark state, corresponding to a strong color
superconductivity with a superfluid gap $\sim 100~$MeV.

In the present paper we performed a general study of  possibility for hybrid
neutron stars with quark cores to reach $M>2~\msun$. We considered a continuum of
parameterized EOS of quark matter, including several existing models. This allowed
us to consider general case of quark cores coexisting with baryonic matter at a
prescribed pressure. We determined  necessary features of baryon - quark-matter
phase transition.  First, the density at which first-order phase transition to
quark phase occurs should be similar to the threshold density for hyperons, $\sim
2\rho_0-3\rho_0$. Second  the relative density jump at the baryon-quark matter
phase transition should be below $30\%$. Third, quark matter should be sufficiently
stiff, which can be expressed as a condition on the sound speed in quark plasma.

The measured $2.0~\msun$ is actually a  lower bound to a true $M_{\rm max}= M^{{\rm
(true)}}_{\rm max}$. The upper bound, resulting the condition of speed of sound
less than $c$ combined with our confidence in the theoretical nucleon EOS for
$\rho<2\rho_0$, is $3.0~\msun$ (see, e.g., \citealt{NSbook2007}, and references
therein). $M^{{\rm (true)}}_{\rm max}$ lies therefore between $2~\msun$ and
$3~\msun$. Obviously, neutron star masses higher than $2.0~\msun$ have to be
contemplated. The question is  how much higher?  The mass of a "black widow" pulsar
could be as high as $2.4~\msun$, albeit the present uncertainty is too large for
this number to be used as an observational constraint (see, e.g.,
\citealt{Lattimer2011}).

To discuss possibility of reaching masses significantly larger than $2.0~\msun$ we
plotted in Fig. \ref{fig:lama_mass} the bounding lines for $M^{\rm (obs)}_{\rm
max}=2.2~\msun$ and $M^{\rm (obs)}_{\rm max}=2.4~\msun$. As we see in Fig.
\ref{fig:lama_mass}, to fulfill condition $M^{\rm (obs)}_{\rm max}=2.4~\msun$ we
have to assume very stiff quark matter, quite close to the  causality limit $a=1$.

The situation becomes even more difficult if we require a strict stability of quark
cores. As a result of high stiffness of quark matter, necessary for $M^{\rm
(obs)}_{\rm max}>2~\msun$ ( and a fortiori for higher lower bounds  $M^{\rm
(obs)}_{\rm max}$), the quark phase turns out to be unstable, beyond some pressure,
with respect to hadronization. Assuming complete thermodynamic equilibrium, we get
very similar $M_{\rm max}$ for the stars with  hyperon and  quark cores.
Consequently, transition to quark matter could not yield $M_{\rm max}>2~\msun$ if
neutron stars with hyperonic cores had $M^{\rm (B)}_{\rm max}$ (significantly) below
$2~\msun$. This is true also for $M^{\rm (obs)}_{\rm max}>2~\msun$. Therefore,
provided our picture of dense matter is valid, we find that a strong hyperon
repulsion at high density is mandatory in general.

The  high-density thermodynamic instability of the quark phase, and its
consequences for $M_{\rm max}$, should be taken with a grain of salt. Our models of
dense baryonic matter assume point  particles. This assumption may be expected to
break down at $\rho\sim 5\div 8\rho_0$. Therefore, the "reconfinement" of the quark
phase is, in our opinion,  likely to  indicate the inadequacy of point-particle
baryonic phase model (see also \citealt{Lastowiecki2012}).

There is another weak point of the commonly used models of quark cores in neutron
stars, characteristic also of the present paper: this is a two-phase approach, with
each phase, baryon and quark one, treated using basically different descriptions.
In principle, both phases and transition between them should have been treated
using a unified approach based on the QCD, so that that the influence of the dense
medium on the baryon structure and baryon interactions are taken into account in a
consistent way. Such approach is beyond the reach of the present day theory of
dense matter. However, a phenomenological modeling of baryon structure in dense
matter is possible, e.g., within a quark-meson coupling model (for references, see
\citealt{Whittenbury2012}). More complete description of neutron-star quark cores,
going beyond the two-phase approximation, should hopefully be obtained  in the
future.

In this paper we were considering non-rotating configurations. Pulsar
PSR J1614-2230 rotates with frequency $f=1/P=317$~Hz and the effect for maximum mass
is of the order of $\simeq 0.01\msun$ \citep{Bednarek2011}, much smaller than
accuracy of mass determination. However it should be noted that for neutron star rotating with
maximum observed frequency of $716$~Hz the effect of rotation would be about five times
larger.

\begin{acknowledgements}
We express our gratitude to David Blaschke and Rafa{\l} {\L}astowiecki for
sharing their numerical results on the sound speed in quark matter. We are grateful
to David Blaschke for reading the manuscript and calling our attention to several
papers relevant to our work. We thank Nicolas Chamel for reading the manuscript
and helpful remarks.
This work was partially supported by the Polish MNiSW research grant no.N N203
512838.
\end{acknowledgements}


\end{document}